\documentclass[nobm,cite,debug]{epl}
\usepackage{epsf}

\title{Orbital and spin Kondo effects in a double quantum dot}
\shorttitle{Orbital and spin Kondo}

\author{Teemu Pohjola\inst{1,2} \and Herbert Schoeller\inst{3} 
        \and Gerd Sch\"on\inst{2,3}}

\institute{
  \inst{1} Materials Physics Laboratory,
        Helsinki University of Technology, Finland\\
  \inst{2} Institut f\"ur Theoretische Festk\"orperphysik,
        Universit\"at Karlsruhe, 76128 Karlsruhe, Germany\\
  \inst{3}      Forschungszentrum Karlsruhe, Institut f\"ur
  Nanotechnologie,  76021 Karlsruhe, Germany
}

\begin{document}

\maketitle

\begin{abstract}

Motivated by recent experiments, in which the Kondo effect has been
observed for the first time in a double quantum-dot structure,
we study electron transport through a system consisting of two ultrasmall, 
capacitively-coupled dots with large level spacing and charging energy.
Due to strong interdot Coulomb correlations, the Kondo effect has two 
possible sources, the spin and orbital degeneracies, and it is maximized 
when both occur simultaneously.
The large number of tunable parameters allows a range of
manipulations of the Kondo physics -- in particular, the Kondo effect 
in each dot is sensitive to changes in the state of the other dot.  
For a thorough account of the system dynamics, the linear and nonlinear 
conductance is calculated in perturbative and non-perturbative
approaches. In addition, the temperature dependence of the resonant peak 
heights is evaluated in the framework of a renormalization group analysis.

\end{abstract}

{\bf Introduction.}
Recently there has been substantial interest in many-body
and correlation effects in ultrasmall semiconductor quantum dots. 
The dots may have strong electron-electron interactions, characterized
by the charging energy  $E_\ab{C}$, and large level spacing $\delta$ 
  (typically $E_\ab{C} > \delta$) \cite{artificial_atoms}.
At low temperature and strong coupling $\Gamma$ 
between the dot and the leads, $k_\ab{B}T\ll\Gamma$, quantum
fluctuations of the charge and  
spin degrees of freedom strongly affect the transport through the dot 
\cite{kondo-theory,Meiretal,diagrams,us,multiorbital}.
The spin fluctuations lead to the Kondo effect, which has been verified in
experiments on single quantum dots\cite{kondo_in_dot}.

\begin{figure}
\epsfxsize=13cm
\centerline{\epsffile{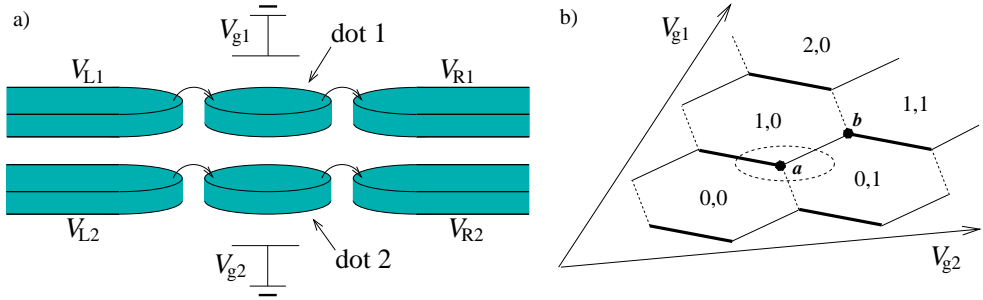}}
\caption{
        {\it a}) Schematic of the capacitively-coupled double quantum dot.
        The bowed arrows denote possibility for tunnelling; there is no 
        direct coupling between the dots.
        {\it b}) Part of the plane spanned by the gate voltages --
        the charge configurations $n_1,n_2$ minimize the charging energy
        within the hexagonal areas. 
        In the Coulomb blockade regime, electron transport is enabled 
        through dot 1 (2) along the thick (dashed) lines, while, in the 
        rest of the plane including the thin solid lines, charge transfer 
        is only possible via cotunnelling and other higher-order processes.
        }
\label{fg:fig1}
\end{figure}

The Kondo effect has recently been discovered also in a double quantum-dot 
structure \cite{MPI}.
Motivated by these experiments, we consider in the present work
a system of two capacitively-coupled quantum dots, depicted in 
Fig.~\ref{fg:fig1}{\it a}). 
The level spacings in each dot are large, 
and effectively just one level per dot is coupled to the two reservoirs.
For strong tunnel coupling and spin-degenerate levels, each dot would 
separately display the usual spin-Kondo effect in its $I-V$ characteristics.
In the present case, the interdot interaction is assumed strong rendering 
the charge states $n_1$ and $n_2$ of the two dots strongly correlated. 
Consequently, the Kondo effect of each dot is sensitive to 
and can be manipulated by voltages applied to the other dot.
The strong interdot interaction has also another, more dramatic effect:
the Kondo effect can arise from the {\it orbital} degeneracy (the two dot 
levels tuned to resonance) even in absence of the real spin degree of 
freedom\cite{MPI}. 
The present model, with two spin-degenerate levels, allows us to study 
the interplay of the two Kondo effects as a function of different voltages
as well as the level splitting.
In absence of orbital degeneracy, the differential conductance can exhibit
satellite resonances for finite bias voltages $V_i$; these may
be easier to observe than split peaks in the case of single dots
({\it e.g.} due to Zeeman splitting \cite{Meiretal,diagrams} or a second 
level in the dot \cite{us}).

%
% preferred position of Fig. 1
%

The paper is organized as follows. 
First the model for the double-dot system is introduced. 
Then we express the current and conductance in terms of the spectral 
density of the dot states in the framework of a real-time diagrammatic 
technique. We calculate the linear and nonlinear conductance through 
each dot as a function of the two transport and the two gate voltages. 
This calculation is complemented by a poor man's scaling 
analysis of the temperature dependence of the linear conductance.

{\bf The model.} 
The double-dot system shown in Fig.~\ref{fg:fig1}{\it a}) is governed by 
the Hamiltonian 
${\mathcal H}={\mathcal H}_\ab{res}+{\mathcal H}_\ab{dot}+{\mathcal H}_\ab{T}$,
where
${\mathcal H}_\ab{res}=\sum_{kri\sigma}
 \epsilon_{kri\sigma}^{}a_{kri\sigma}^\dagger a_{kri\sigma}^{}$ 
describes non-interacting electrons in the reservoir $ri$
($r=\ab{L,R}$ and $i=1,2$), 
${\mathcal H}_\ab{dot}=\sum_{i\sigma} 
  \xi_{i\sigma}^{} c_{i\sigma}^\dagger c_{i\sigma}^{} + U_{n_1,n_2}$ 
refers to the dot electrons, and 
${\mathcal H}_\ab{T}=\sum_{kri\sigma} 
  (T_{ri}^{k\sigma} a^\dagger_{kri\sigma} c_{i\sigma}^{}\,+\,H.c.)$ 
is the tunnelling Hamiltonian. For each barrier we introduce the
 tunnelling strength $\Gamma_{ri}=2\pi\sum_k|T_{ri}^{k\sigma}|^2\delta
 (\omega-\varepsilon_{kri\sigma})\approx2\pi|T_{ri}|^2\rho_0$, 
which is assumed to be independent of energy in the range 
of interest ($\rho_0$ is the density of states in the leads).
The electron-electron interaction, for $n_i$ electrons in the dot $i$,
is accounted for by the charging energy  $U_{n_1,n_2}$.
In general, the interaction depends on all the voltages applied to the system,
but for symmetrically applied transport voltages it is determined by
the gate voltages $V_{\ab{g}1}$  and $V_{\ab{g}2}$ alone.

In the plane spanned by these gate voltages, within the hexagonal
areas depicted in Fig.~\ref{fg:fig1}{\it b}), 
the ground state is reached for the charge configurations $(n_1,n_2)$ 
indicated in the figure, and electron tunnelling is suppressed by Coulomb
blockade effects. 
In the figure, there are three kinds of boundaries in the honeycomb lattice. 
Electron transport through dot 1 (2) is only possible 
along the boundaries indicated with the thick (dashed) lines, while
the thin lines correspond to degeneracy between the two dots, in particular, 
$\xi_{1\sigma}=\xi_{2\sigma}$ along the $a$-$b$ line.
As in the experiment, where the two dots lie close on top of each other, 
both the intradot and interdot interactions are assumed to be strong.
Hence, for low temperature and bias voltages, 
$k_\ab{B}T,eV_i\ll E_\ab{C}$, it suffices to consider three charge states.
Below we focus on the circled region in Fig.~\ref{fg:fig1}{\it b}), {\it i.e.} 
the states (0,0), (0,1), and (1,0).
In this restricted basis, the states can be newly defined to include
the charging energy $U_{n_1,n_2}$, {\it e.g.}, for an electron 
with spin $\sigma$ in dot $1$, 
$\xi_{1\sigma}+U_{1,0}-U_{0,0}\rightarrow\varepsilon_{1\sigma}$.
We observe that the resulting Hamiltonian ${\mathcal H}$ has a one-to-one 
correspondence to a single-dot model with two (spin-degenerate) orbital 
states, distinguished by an orbital quantum number $i$.
The index $i$ corresponds to the spatially separated dots and is therefore 
conserved in the tunnelling processes.

{\bf Transport.}
We aim to evaluate the differential conductance 
$G_i(V_1,V_2)=\partial I_i/\partial V_i$ ($G\equiv G_1+G_2$) from the current 
through the dot $i$. The dc-current can be expressed as \cite{Meiretal}
\begin{eqnarray}
  I_i = e \sum_\sigma 
      \frac{\Gamma_{\ab{L}i}\Gamma_{\ab{R}i}}
           {\Gamma_{\ab{L}i}+\Gamma_{\ab{R}i}}
      \int_{-\infty}^\infty \upd\omega         
        [f_{\ab{R}i}(\omega)-f_{\ab{L}i}(\omega)]A_{i\sigma}(\omega).
\label{eq:current}
\end{eqnarray}
Here $f_{ri}(\omega)$ are the Fermi functions of the leads with
chemical potentials $\mu_{ri}$.
The spectral density 
$A_{i\sigma}(\omega)\equiv 
  [C_{i\sigma}^<(\omega)-C_{i\sigma}^>(\omega)]/2\pi i$ 
of the state $i\sigma$ corresponds to a local density of states in 
the quantum dot(s). Its shape is reflected in the differential conductance.

\begin{figure}
\epsfxsize=13.5cm
\centerline{\epsffile{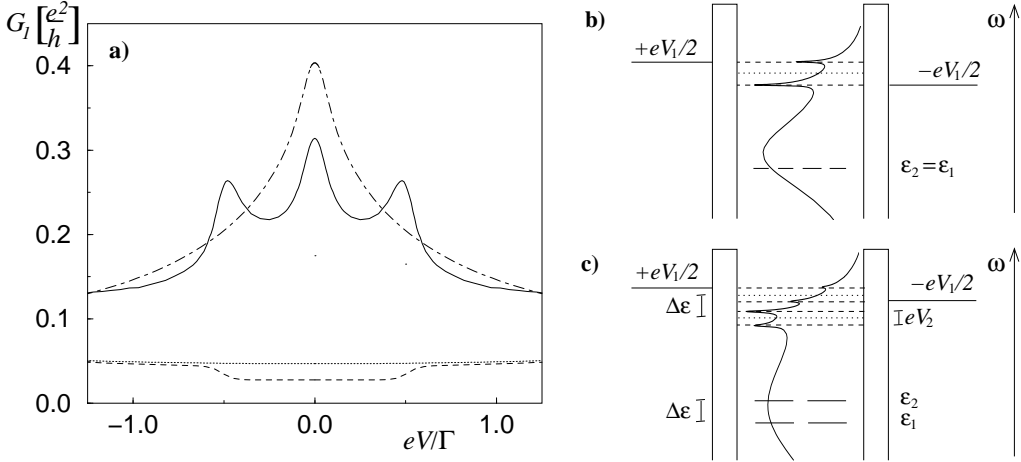}}
\caption{
  {\it a}) Differential conductance $G_1$ through dot 1 as a function of
  the bias voltage $V=V_1=V_2$ and for a fixed 
  $\varepsilon_{1\sigma}=-4\Gamma$. Temperature $k_\ab{B}T=\Gamma/50$.
  The two lower curves are the cotunnelling contributions to the conductance
  for $\Delta\varepsilon=0$
  (dotted) and $\Delta\varepsilon=\Gamma/2$ (dashed).
  The higher curves are results in the resonant-tunnelling approximation
  the same sets of parameters (dot-dashed and solid, respectively)
  displaying zero-bias maxima as well as two satellites for 
  $\Delta\varepsilon\neq0$.
  {\it b}) and {\it c}) show $A_{1\sigma}(\omega)$ between two tunnel 
  barriers; also the chemical potentials at $\mu_{r1}=\pm eV_1/2$ are 
  indicated. In {\it b}) $\Delta\varepsilon=0$ and $V_1=V_2=V$, while in 
  {\it c}) $\Delta\varepsilon\neq0$ (still $V_1=V_2$).
  The four solid lines denote the energies of the states $1\uparrow$, 
  $1\downarrow$, $2\uparrow$, and $2\downarrow$.
  The states at $\varepsilon_{1\sigma}$ give rise to the broad 
  Lorentzian shape in $A_{1\sigma}(\omega)$. 
  The positions of the sharp Kondo resonances in $A_{1\sigma}(\omega)$
  are depicted as the dashed lines (dotted lines denote the case $V_i=0$),
  and the other relevant energies are shown on the sides.
        }
\label{fg:fig2}
\end{figure}

In the weak-coupling limit, the lowest-order perturbation theory 
yields $A_{i\sigma}^{(0)}(\omega)\propto\delta(\omega-\varepsilon_{i\sigma})$ 
which describes the suppression of the current, {\it i.e.}, Coulomb blockade 
effects.
In second order, or `cotunnelling' regime, the spectral density takes the form 
\begin{eqnarray}
\label{eq:A_2}
    &&A_{i\sigma}^{(1)}(\omega) =
        \ab{Re}\frac{1}{(\omega-\varepsilon_{i\sigma}+i0^+)^2}
        \times
        \Big\{ (p_0^{(0)} + p_{i\sigma}^{(0)})
                 (\gamma_i^+(\omega)+\gamma_i^-(\omega))\\
    &&   \;\;\;\;\;\;\;\;\;\;\;\;\;\;\;\;\;\;\;\;\;\;\;\;\;\;\;\;\;\;
        \;\;\;\;\;\;\;\;\;\;\;\;\;\;\;\;\;\;
        +\sum_{j\sigma^\prime\neq i\sigma}
        \big[p_{i\sigma}^{(0)}\gamma_j^+(\omega-\varepsilon_{i\sigma}
                                +\varepsilon_{j\sigma^\prime})
    + p_{j\sigma^\prime}^{(0)}\gamma_j^-(\omega-\varepsilon_{i\sigma}
                                        +\varepsilon_{j\sigma^\prime})
     \big]\Big\},\nonumber
\end{eqnarray}
where $p_{i\sigma}^{(0)}$ are the occupation probabilities for the states
$i\sigma$ as obtained in the lowest-order perturbation theory
\cite{detailed_balance}
and $2\pi\gamma_i^\pm(\omega)=\sum_r\Gamma_{ri}f_{ri}^\pm(\omega)$.
As the prefactor in Eq.~(\ref{eq:A_2}) suggests, $A_{i\sigma}(\omega)$ only 
decays algebraically as $1/(\omega-\varepsilon_{i\sigma})^2$
leading to a non-vanishing conductance everywhere in the plane in 
Fig.~\ref{fg:fig1}{\it b}).
The resulting linear conductance -- denoted as $g$ in the following --
as a function of 
$\Delta\varepsilon\equiv\varepsilon_{2\sigma}-\varepsilon_{1\sigma}$
shows a maximum for $\Delta\varepsilon=0$, {\it i.e.}, on the line $a$-$b$ 
in Fig.~\ref{fg:fig1}{\it b}) (this straight-forward result is not displayed).
As a function of the transport voltage $V$, for symmetric applied 
voltages $V=V_{i}=V_{\ab{R}i}-V_{\ab{L}i}$ for both $i$, 
the differential conductance exhibits smooth variations, 
which are displayed as the two lowest curves in Fig.~\ref{fg:fig2}{\it a}).

For stronger coupling,  further higher-order processes become important.
We account for them within the so-called 
resonant-tunnelling approximation \cite{diagrams}, which amounts to a
resummation of a certain class of diagrams (with maximally two-fold
off-diagonal density matrix) in all orders in the tunnelling amplitudes. 
To gain some insight into the results, let us consider again
the symmetric applied voltages, equal barriers, $\Gamma_{ri}=\Gamma_r$, and 
degenerate levels $\varepsilon_{i\sigma}=\varepsilon_0$
({\it i.e.} the $a$-$b$ line in Fig.~\ref{fg:fig1}{\it b})). 
In this case, all the spectral densities are equal,
$A_{i\sigma}(\omega)=A(\omega)$, and can be calculated explicitly:
\begin{eqnarray}
\label{eq:Juergens_A}  
  A(\omega)=\frac{1}{\pi}\frac{\Gamma/2}
   {[\omega-\varepsilon(\omega)]^2 + [\sum_r(\Gamma_r/2)\{1+(N-1)
        f_r(\omega)\}]^2}.
\end{eqnarray}
Here $N$ denotes the degeneracy of the resonance and the level renormalization
$\varepsilon(\omega)=\varepsilon_0+(N-1)
\sum_r(\Gamma_r/2\pi)\ab{ln}[E_\ab{C}/
\ab{max}\{2\pi k_\ab{B}T,eV/2,|\omega|\}]$. 
For a single level without spin, $N=1$, 
only charge fluctuations are possible 
and Eq.~(\ref{eq:Juergens_A}) reduces to the Breit-Wigner formula with 
an unrenormalized Lorentzian form.
For two spinless levels, $N=2$, and the orbital index $i$ 
behaves like a pseudospin; for two spin-degenerate levels we have $N=4$.
If $N\ge2$, the (pseudo)spin dynamics contributes to 
Eq.~(\ref{eq:Juergens_A}) and, at low temperature
and for small excitation energies, it dominates over the charge fluctuations. 
As a consequence an additional sharp resonance, the Kondo peak, emerges in 
$A(\omega)$ at the positions of the chemical potentials in the leads. 
Figure \ref{fg:fig2}{\it b}) 
shows an example of $A(\omega)$ for $N=4$ and introduces
a schematic way for visualizing its relevant structure.
The Kondo peak is reflected in the conductance as a sharp zero-bias maximum, 
dot-dashed curve in Fig.~\ref{fg:fig2}{\it a}), which is a distinctive sign of 
the Kondo effect. The peak in $G$ is the higher the larger $N$ is,
in accordance with the $N$-dependent Kondo temperature, see below.

%
% Preferred position of Fig. 2
%

The most intriguing results are found for the general case with 
non-degenerate levels 
($\varepsilon_{2\sigma}\neq\varepsilon_{1\sigma}$ for detuned dots,
and/or $\varepsilon_{i\uparrow}\neq\varepsilon_{i\downarrow}$ in  
a magnetic field) and/or different bias voltages, $V_1\neq V_2$.
In order to account for these extensions,
we calculate $A_{i\sigma}(\omega)$ numerically.
For brevity of the following discussion, we set the magnetic field 
to zero and focus on the conductance $G_1$ through dot 1.
Figure \ref{fg:fig2}{\it a}) (solid line) shows an example of 
$G_1$ for equal bias voltages, $V_i=V$, and a finite level separation 
$\Delta\varepsilon$.
The conductance reflects the shape of the spectral function 
$A_{1\sigma}(\omega)$, illustrated in Fig.~\ref{fg:fig2}{\it c}), which 
exhibits resonances at $\pm eV_1/2$ and $-\Delta\varepsilon\pm eV_2/2$.
The former arise due to the spin fluctuations in dot 1 and
are similar to the usual Kondo peaks; the latter are due to tunnelling 
processes involving both levels -- {\it i.e.} fluctuations of the orbital 
index $i$ -- and thus depend on the level separation $\Delta\varepsilon$.
As a rule, the conductance $G_1$ shows a peak whenever two of the resonances 
in $A_{1\sigma}(\omega)$ coincide:
there is a zero-bias maximum due to the spin-degeneracy of the levels
(although the levels are now detuned) and,
in addition, two satellites at $eV=\pm\Delta\varepsilon$ --
these correspond to the additional peaks in $A_{1\sigma}(\omega)$
and reflect the presence of the other level/dot.

An experimental observation of the satellites in $G_1$ is facilitated by 
two points.
First, finite bias voltages induce decoherence to the electron transport and 
tend to suppress the Kondo effect for too high voltages \cite{Kaminskietal}.
However, in the present case the splitting $\Delta\varepsilon$ can be 
controlled starting from zero and only small bias voltages are required
in observing the splitting.
Second, due to the larger number of states (=4) the split peaks are more
pronounced in the present system than, {\it e.g.},  for a single dot with one 
Zeeman-split level (altough they have been observed \cite{kondo_in_dot}
in this case as well).

\begin{figure}
\epsfxsize=14cm
\centerline{\epsffile{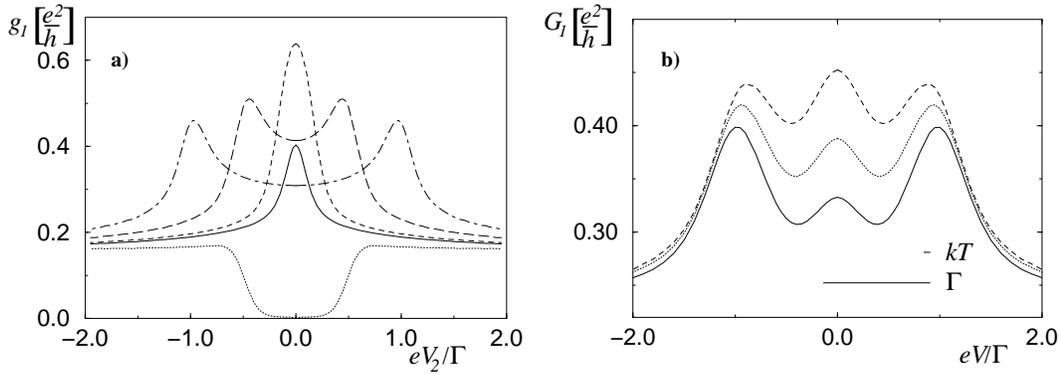}}
\caption{
  {\it a}) Linear conductance $g_1$ as a function of $eV_2$ for
  $\varepsilon_{1\sigma}=-4\Gamma$ and  
  $\varepsilon_{2\sigma}=-4.25\Gamma$ (dotted), $-4\Gamma$ (solid),
  $-3.94\Gamma$ (dashed)\cite{peak_max}, $-3.75\Gamma$ (long dashed),
  and $-3.5\Gamma$ (dot-dashed).
  {\it b}) Differential conductance $G_1$ at $k_{\rm B}T=\Gamma/10$ and 
  the levels tuned to the mixed-valence regime -- the {\it renormalized} 
  upper level crosses the Fermi energies.
  The bare levels are at $\varepsilon_{1\sigma}=-3.5\Gamma$ (solid),
  $-3.375\Gamma$ (dotted), and $-3.25\Gamma$ (dashed) and 
  $\Delta\varepsilon=\Gamma$;
  the renormalization is roughly $2.4\Gamma$ upwards.
  The energy scales $\Gamma$ and $k_{\rm B}T$ are also shown.
        }
\label{fg:fig3}
\end{figure}

The fact that the voltages $V_i$ can be chosen independently opens new
interesting possibilities.
Let us consider the spectral density in Fig.~\ref{fg:fig2}{\it b}) 
and the splitting of the Kondo resonance by the applied bias voltage. 
In the case of single quantum dots, the voltage-induced split in $A(\omega)$
is reflected as a mere monotonous decrease of the differential conductance $G$.
In the present system, however, the {\it linear} conductance $g_1$ is
found to show a peaked structure as a function of the voltages $V_2$ and 
$V_{\ab{g}2}$ ($\sim\varepsilon_{2\sigma}$) applied to the {\it other} dot, 
see Fig.~\ref{fg:fig3}{\it a}). The peaks in $g_1$ occur when the condition 
$eV_2=\pm2\Delta\varepsilon$ is fulfilled, see Fig.~\ref{fg:fig2}{\it c}) and above.

Before embarking on the scaling analysis, let us make a remark
concerning the ``Kondo regime'':
it turns out that $G_1$ may display a peaked structure also
outside this regime.
Let us consider a higher temperature than above, {\it e.g.} 
$k_{\rm B}T=\Gamma/10$, and 
tune the system into the mixed-valence regime with the levels brought into
a close vicinity of the Fermi energies in the leads. 
In this case, no Kondo effect is expected.
A perturbative calculation produces the usual Coulomb blockade peaks, which
in this case are four peaks that shift with $\varepsilon_{i\sigma}$;
the non-perturbative result for $G_1$ shown in Fig.~\ref{fg:fig3}{\it b})
displays a triple-peak structure with one peak at zero-bias and two satellites 
at $eV=\pm\Delta\varepsilon$.
The number and position of these peaks suggest that the peaks in $G_1$ could 
be due to the Kondo effect after all, cf. results obtained above.
On the other hand, if the peaks were of the Kondo origin, their width 
should reflect $\ab{max}\{k_\ab{B}T,k_\ab{B}T_\ab{K}\}$
(or $\tau^{-1}$, the inverse decoherence time of Ref.~\cite{Kaminskietal},
should this exceed the temperature), 
where $T_\ab{K}$ is the new low-energy scale, the Kondo temperature.
However, in Fig.~\ref{fg:fig3}{\it b}) the resonance widths are seen 
to be almost $\Gamma$, see the energy scales indicated in the figure;
see also Ref.~\cite{mixed-valence}.
Hence, although the peak widths are not a strict measure of the Kondo
regime, we conclude that higher-order tunnelling processes may also 
give rise to non-trivial peaked structure other than the Kondo type.

%
% Preferred position of Fig. 3
%

%%%%%%%%%%%%%%%%%%%%%%%%%%%%%%%%%%%%%%%%%%%%%%%%%%%%%%%%%%%%%% scaling %%%%

{\bf Scaling.}
In this section, we complement the above considerations in the framework 
of the poor man's scaling technique \cite{Hewson}, 
a renormalization group approach 
suited for the Kondo regime at equilibrium (zero-bias limit).
This approach is used to derive explicit expressions for 
the logarithmic temperature dependence of the linear conductance $g$ --
experimentally the most convincing sign of the Kondo effect --
and the Kondo temperature $T_\ab{K}$.

In order to obtain the proper scaling form of the conductance,
we consider the dots deep in the Coulomb blockade regime,
$\varepsilon_{i\sigma}\ll\mu_{ri}=0$, and start by Schrieffer-Wolff (SW) 
transforming the Hamiltonian ${\mathcal H}$ into the scattering form
\begin{eqnarray}
  {\mathcal H}_{\rm scat} &=&
    \sum_{r,r^\prime}\sum_{k,k^\prime}\sum_{i\sigma,j\sigma^\prime}
    \left[
      J_{rk,r^\prime k^\prime}^{i\sigma,j\sigma^\prime}
        c_{rki\sigma}^\dagger c_{r^\prime k^\prime j\sigma^\prime}^{}
        c_{j\sigma^\prime}^\dagger c_{i\sigma}^{}
    - \frac{1}{N}
      J_{rk,r^\prime k^\prime}^{j\sigma^\prime,j\sigma^\prime}
        c_{rkj\sigma^\prime}^\dagger c_{r^\prime k^\prime j\sigma^\prime}^{}
        c_{i\sigma}^\dagger c_{i\sigma}^{}
    \right],\\
\label{eq:scattering_Hamiltonian}
  J_{rk,r^\prime k^\prime}^{i\sigma,j\sigma^\prime}
      &=& \frac{1}{2}T_{ri}^{k\sigma}(T_{r^\prime j}^{k^\prime\sigma^\prime})^*
          \left(\frac{1}{\varepsilon_{i\sigma}+E_\ab{C}}-
            \frac{1}{\varepsilon_{i\sigma}}
          +\frac{1}{\varepsilon_{j\sigma^\prime}+E_\ab{C}}
           -\frac{1}{\varepsilon_{j\sigma^\prime}}\right).
\end{eqnarray}
For $N$ degenerate states, $\Delta\varepsilon=0$, ${\mathcal H}_\ab{scat}$ 
corresponds to the Coqblin-Schrieffer model\cite{Hewson}.
The current through both dots, $I=I_1+I_2$, calculated in the lowest 
(second) order in $J$ is
\begin{eqnarray}
\label{eq:I_transformed}
  I=2\pi e\sum_{i\sigma}\sum_{j\sigma^\prime}
   \{p_{i\sigma}^{(0)}\alpha_{\ab{R}j\rightarrow\ab{L}i}^{}
        (\varepsilon_{j\sigma^\prime}-\varepsilon_{i\sigma})
        - p_{j\sigma^\prime}^{(0)}\alpha_{\ab{L}i\rightarrow\ab{R}j}^{}
        (\varepsilon_{i\sigma}-\varepsilon_{j\sigma^\prime})\},
\end{eqnarray}
where $p_{i\sigma}^{(0)}$ are the same probabilities that
entered Eq.~(\ref{eq:A_2}) and 
\begin{eqnarray}
\label{eq:rate_scaling}
  \alpha_{ri\rightarrow r^\prime j}(\omega)
        = (J_{ri,r^\prime j}\rho_0)^2
        \frac{\omega-\mu_{ri}+\mu_{r^\prime j}}
             {\exp[\beta(\omega-\mu_{ri}+\mu_{r^\prime j})]-1}
\end{eqnarray}
are the tunnelling rates from reservoir $ri$ to $r^\prime j$
\cite{form_of_rates}.
Here $T_{ri}^{k\sigma}=T_{ri}$ is assumed, implying $J_{ri,r^\prime
j}^{}=J_{rk,r^\prime k^\prime}^{i\sigma,j\sigma^\prime}$.
It should be noted that despite its different appearance,
the present formulation yields just the cotunnelling current expressed above
in terms of Eqs.~(\ref{eq:current}) and (\ref{eq:A_2}).
However, Eqs.~(\ref{eq:I_transformed}) and (\ref{eq:rate_scaling}) are more 
suitable for the scaling analysis to follow.

In the next step towards $g(T)$,
we perform poor man's scaling to second order in the coupling constants.
In the degenerate case, $\Delta\varepsilon=0$, there is just one coupling 
constant $J$ in the problem and it is renormalized as 
$\tilde{J}=J[1+N\rho_0J\ln\frac{\tilde{D}}{D}]^{-1}
 =[N\rho_0\ln\frac{\tilde{D}}{k_\ab{B}T_\ab{K}}]^{-1}$
\cite{nondeg_scaling}.
The initial high-energy cutoff $D$ corresponds to the band 
width in the leads, while the final cutoff is taken as
$\tilde{D}=\max\{k_\ab{B}T,eV\}$ 
(or $\tilde{D}=\tau^{-1}$, the decoherence rate of Ref.~\cite{Kaminskietal}).
The second equality for $\tilde{J}$ follows from the definition of the Kondo 
temperature $k_\ab{B}T_\ab{K} = D\exp[-1/N\rho_0J]$.
This in turn yields an increasing $T_\ab{K}$ for increasing $N$ and
gives a qualitative reason, why the zero-bias peak in $G_1$ is 
emphasized for larger values of $N$, see above.

Assuming that the zero-bias limit can be modelled in terms of equilibrium
properties of the system, we insert the scaled coupling constants into 
Eq.~(\ref{eq:I_transformed}) and obtain
\begin{eqnarray}
\label{eq:scaling_of_g}
  g(\tilde{D}) = \frac{2\pi e^2}{\hbar}\frac{N^2-1}{N^3}
        \ab{ln}^{-2}\left(\frac{\tilde{D}}{k_\ab{B}T_\ab{K}}\right)
\end{eqnarray}
for the leading cutoff dependence of the peak conductance.
The conductance $g_1(T)$ through just one dot 
is sensitive to the voltage $V_2$:
If $V_1=V_2=V$, $g_1(T)=g(T)/2$, while for $V_2\equiv0$, 
$g_1(T)$ is obtained from Eq.~(\ref{eq:scaling_of_g}) by replacing
the $N$-dependent part by $(N^2+4N-2)/4N^3$.

{\bf Summary.}
In conclusion, we have studied electron transport through 
a capacitively-coupled double quantum dot with a large level 
spacing and intra and interdot charging energies.
The system is found to exhibit the Kondo effect due to fluctuations in both 
the orbital and spin degrees of freedom. 
The linear and nonlinear conductance is calculated for 
arbitrary level positions and bias voltages and it is found 
to display a rich structure both in and out of the Kondo regime.
For detuned levels, the differential conductance displays zero-bias
as well as satellite peaks. The side peaks are argued to be 
experimentally observable despite the suppressive effect of finite bias 
voltages~\cite{Kaminskietal}.
The Kondo effect in each dot is shown to be very sensitive to the voltages 
applied to the other dot allowing delicate manipulation of the Kondo physics.
An example of voltage-induced splitting of the Kondo peak has been discussed.

\acknowledgments

We would like to acknowledge J.~Schmid, J.~Weis, and U.~Wilhelm 
for discussions.
This work has been supported by the Finnish Cultural Foundation,
EU TMR network ``Dynamics of Nanostructures'',
the Swiss National Foundation, and DFG through SFB 195.


\begin{thebibliography}{16}
\bibitem{peak_max}
The peak is maximized for a finite level splitting,
$\varepsilon_{1\sigma}<\varepsilon_{2\sigma}$,
due to the increase in the occupation probability of dot 1 and
the subsequent increase in the weight under $A_{1\sigma}(\omega)$.
\end{thebibliography}
\end{document}